\documentstyle[12pt]{article} 
\include{./psfig}
\begin{document}

\title{Adaptive Detection of Instabilities: An Experimental
Feasibility Study.}
\author{R. Rico-Mart\'{\i}nez$^{1,2}$\thanks{
Corresponding author. E-mail: ramiro@arnold.princeton.edu, Telephone
+52-46117802, Fax +52-46117744},
K. Krischer$^2$, G. Fl\"atgen$^2$, \and J.S. Anderson$^3$ and
I.G. Kevrekidis$^{2,3}$\\
$^1$ Instituto Tecnol\'ogico de Celaya, Depto. de Ingenier\'{\i}a 
Qu\'{\i}mica,\\ Celaya, Gto. 38010 M\'exico \\
$^2$ Fritz-Haber-Institut der Max-Planck-Gesellschaft,\\
Faradayweg 4-6, 14195 Berlin, Germany\\
$^3$ Department of Chemical Engineering, Princeton University, 
\\Princeton, NJ 08544 U.S.A.\\}

\maketitle

\begin{abstract}

We present an example of the practical implementation of a
protocol for experimental
bifurcation detection based on {\it on-line}
identification and feedback control ideas.
The current experimental practice for the detection
of bifurcations involves a cumbersome
approach typically requiring long observation times in the
vicinity of marginally stable solutions, as well as
frequent re-settings of the experiment for the
detection of turning-point or subcritical bifurcations.
The approach exemplified here addresses these issues drawing from
ideas of numerical bifurcation detection.
The idea is to couple the experiment with an on-line computer-assisted
identification/feedback protocol so that the closed-loop
system will converge to the open-loop bifurcation points.
We demonstrate the applicability of
this instability detection method by real-time, computer-assisted
detection of period doubling bifurcations of an electronic circuit;
the circuit implements an analog realization of
the R\"ossler system.
The method succeeds in locating the bifurcation points even
in the presence of modest experimental uncertainties, noise and limited
resolution.
The results presented here include bifurcation detection
experiments that rely on measurements
of a single state variable and delay-based phase space reconstruction, as
well as an example of tracing entire segments of a codimension-1
bifurcation boundary in two parameter space.

\leftline{{\bf PACS numbers: 05.45.,82.40}}

\leftline{{\bf Keywords:} Bifurcation Detection, Nonlinear Systems, Adaptive
Control.}
\end{abstract}
\newpage

\section{{\bf Introduction}}
\thispagestyle{empty}
\noindent
Over the last 30 years increasing efforts have been directed towards
the understanding of experimentally observed complex dynamical behavior.
The development of Nonlinear Dynamics as a branch of applied 
mathematics provides us the set of tools 
necessary for understanding the response
of nonlinear dynamical systems 
as operating parameters are varied.
Characterization of the nonlinear
dynamical behavior of experimental systems, however, remains a difficult task.
This is mainly because real systems cannot be manipulated as 
easily or as cleanly as a model would on
the computer when its bifurcation behavior is traced.

Typically, an  experimental bifurcation diagram is
obtained via repeated application of a {\it set-and-observe}
procedure. The operating parameter of interest
is set and the system dynamics allowed to evolve asymptotically to stationary 
(e.g. steady or oscillatory) behavior.  
The operating parameter setting is then changed (e.g. incremented slightly)
and the experiment is repeated; if a qualitative difference is detected
between the asymptotic dynamics reached for two consecutive parameter
settings, a bifurcation is inferred to have occurred somewhere in-between.
A similar approach  can be applied to the {\it computational} detection
of bifurcations if one 
is only allowed to use forward-in-time integration of the model equations 
(i.e. set parameter values and initial conditions and run a simulation to
stationarity).
In both cases there are serious drawbacks.
For example, in the investigation of a period doubling bifurcation
of a periodic orbit, as the 
critical value of the search parameter is approached the dynamics will
slow down considerably, and the time required to determine whether ultimately 
single loop or double loop oscillations are
observed increases dramatically.
Furthermore, if one wishes to detect a subcritical bifurcation, or
a saddle-node bifurcation, once one steps over the critical parameter
value, the
system will move far away from the phase space region of interest.
The experiment will then need to be reset to reach the vicinity of the
stable solution located just before the critical parameter value
that was overstepped. 

In previous work \cite{ciss,prl} we proposed a protocol for the
experimental detection of bifurcations motivated by the procedure
of computational construction of bifurcation diagrams from mathematical
models.
Numerical bifurcation theory in a computational
environment allows the bifurcation location via a simultaneous
search for criticality in (phase)$\times$(parameter) space.
An augmented set of equations (e.g. steady state equations 
plus a criticality condition) is typically formed, and its fixed point 
(steady state values at criticality {\it and} critical
parameter value) is converged upon quadratically using a Newton-Raphson
contraction mapping for the augmented system of equations.
Obviously in this case the states do not evolve on the computer
based on the system
dynamics, but based on the computational contraction mapping dynamics.

The proposed protocol for the detection of bifurcations seeks an
``experimental-computational" compromise:
an experiment (i.e. states evolving based on system dynamics) is coupled
to a parameter evolution law which is based on criticality conditions and
feedback control. The augmented, closed loop system,
constructed in this
manner has a {\it stable} steady state at what used to be
the open loop system's {\it marginally stable} steady state.
The computational building blocks that
constitute the parameter evolution law are based on  the
{\it on-line} identification of
a ``local nonlinear" dynamic model, obtained via least-squares regression
\cite{PTVF},
used to estimate criticality on-line, and simple feedback control
algorithms that use the bifurcation parameter as the actuator in order to
drive the system towards the bifurcation point.
As it will be illustrated below,
this algorithm enhances our ability
to efficiently detect instabilities, trace segments of the
bifurcation diagram (``operating diagram"),
and, in general, characterize the system dynamics in multiparameter space.
The ``experimental" system used in our illustration is an analog realization
of the R\"ossler system. For this system, we converge experimentally
on the period doubling
bifurcation of simple periodic orbits in one and two-parameters. The
local adaptive model used to infer the presence of the bifurcation is a 2D nonlinear
map. Our illustrations are based on (a) full state vector measurement; and 
(b) reconstruction of the state-space based on time series of a single state 
measurement and the use of delay coordinates.

An alternative motivation for studying this type of problem is the
typical Ziegler-Nichols (ZN) tuning procedure for PID (proportional-
integral-derivative) controllers
\cite{ziegler},
which motivated the work of O'Neil, Lyberatos and Svoronos
\cite{Lyb} on adaptive determination of bifurcation points.
In the ``textbook" ZN procedure the ``experiment" is the 
closed loop system with a proportional controller; the gain of
this controller is the ``bifurcation parameter".
The closed-loop system loses stability when the proportional gain $K$
reaches the value $K_u$. 
This critical parameter value is precisely the ZN ``ultimate gain".
In practice, one obtains $K_u$ by slowly increasing the gain of 
the proportional controller
until one observes incipient oscillations.
Knowledge of the location of the bifurcation and the frequency of the oscillations
dictates then the ZN-tuning settings for a PID controller;
it is interesting that this practical controller tuning approach is based
on experimental bifurcation detection.
%

The paper is organized as follows: first, we describe
our system (the electronic R\"ossler circuit) and the bifurcation 
detection protocol before presenting
our experimental results. We then provide some concluding remarks. 

\begin{figure}[t]
\centerline{\psfig{file=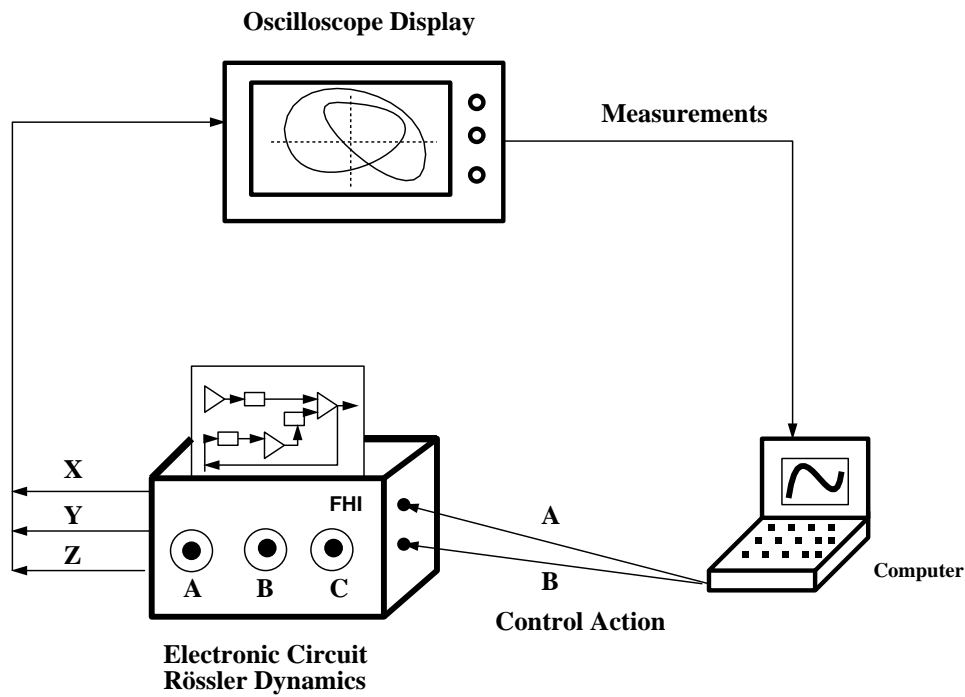,width=5.0in}}
\caption{Schematic of the experimental setup.}
\label{schem}
\end{figure}

\section{{\bf The R\"ossler Circuit}}

The R\"ossler model was proposed in the early 1970's \cite{rossler}
as one of ``the simplest" nonlinear vectorfields 
capable of generating chaotic behavior.
It consists of the following set of three ordinary differential equations:

\begin{eqnarray}
{{dx}\over{dt}} & = & -y -x \\
{{dy}\over{dt}} & = & x + ay \\
{{dz}\over{dt}} & = & b + z(x-c) 
\end{eqnarray}

This system has been extensively studied. For $b=2$ and $c=4$ and as $a$
is increased, it exhibits a steady state
undergoing a Hopf bifurcation followed by a cascade of period doubling
bifurcations of the periodic orbits culminating in an apparently chaotic 
attractor around $a=0.396$.

\begin{figure}[t]
\centerline{\psfig{file=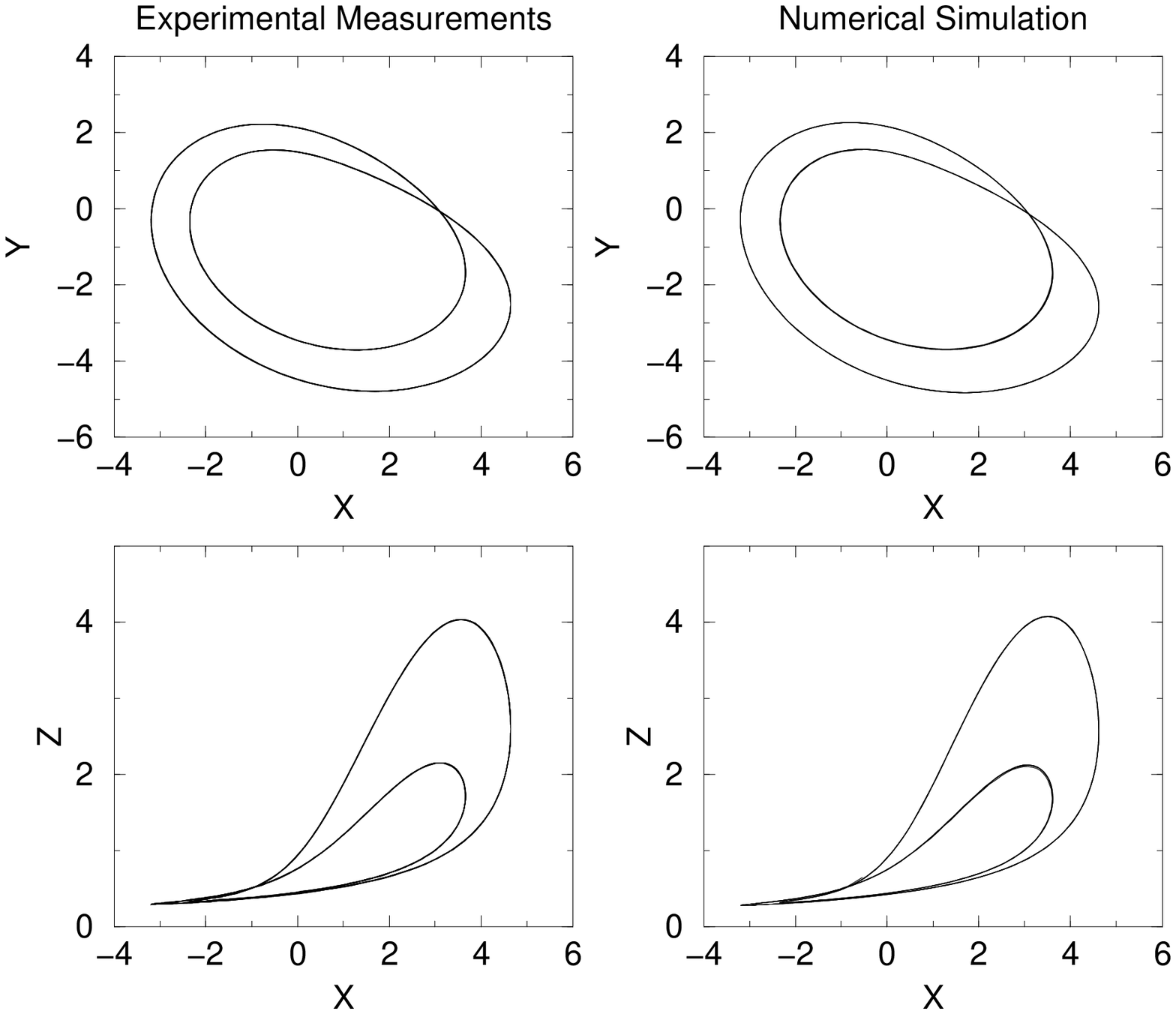,width=5.0in}}
\caption{Comparison of attractors for the circuit (left) and the
numerically computed trajectory obtained integrating the set of ODEs
of the R\"ossler system. The comparison is made for a double-loop
trajectory observed at $a=0.35$, $b=2$ and $c=4$.}
\label{rosscomp}
\end{figure}

In the implementation of the protocol for the adaptive detection of
experimental bifurcations, we use as model system an analog realization
of the R\"ossler system. The R\"ossler equations
are implemented via a sequence of feedback loops of electronic signals.
The circuit was manufactured at the FHI labs\footnote{
A detailed diagram of the circuit is available in the web in
the following address
http://w3.rz-berlin.mpg.de/pc/spatdyn/spatdyn.cgi?Publications}
and includes an encasing
that allows both manual and digital manipulation of all three system parameters. 
In our experiments the parameter dynamics were computer-generated  in closed
loop, and were communicated to the device through its digital I/O channels. 
The circuit is coupled (Fig.~\ref{schem})
with a VMEbus System (32bit backplane system) consisting of a
VMEbus CPU-board Baja40 from Heurikon Division controlled
via a Motorola 68040 32-bit microprocessor chip running internally
with 66 MHz, externally with 33 MHz, 32 MByte RAM and
several I/O interfaces and a VMEbus IP carrier board with
four IndustryPack slots. One slot was used
with a TIP501 optically isolated 16 channel
16 bit ADC module from TEWS Datentechnik, input voltage +/- 10 volt,
14.5 usec data acquisition and conversion time,
accuracy +/- 4 LSB, linearity +/- 4 LSB, and the other slot with a
TIP550 optically isolated 4 channel 12 bit DAC from TEWS
Datentechnik, output voltage +/- 10 volt, 13 usec settling time,
accuracy +/- 1 LSB, linearity +/- 1 LSB. The array uses an
operating system VxWorks 5.3.1 (Tornado) from Wind River System.
The data acquisition
components allow us to specify the accuracy of the measurements from
11 to 16 bits.

The dynamic response of the motivating set of differential equations
is well reproduced. Figure~\ref{rosscomp} presents a comparison of a couple
of projections of the double-loop limit cycle predicted and observed at $a=0.35$,
$b=2$ and $c=4$. The deviation of the signal given by the circuit with respect
to the one obtained by integrating the differential equations is minimal,
both in the actual values of the variables as well as in the period of the
oscillations. Similar results are observed within the range of operation
of the circuit: $a \in [0,1]$, $b\in [0,10]$, $c\in [0,10]$ and values of
the state variables inside the range [-10,10].

\section{{\bf The Bifurcation Detection Protocol}}

\begin{figure}[t]
\centerline{\psfig{file=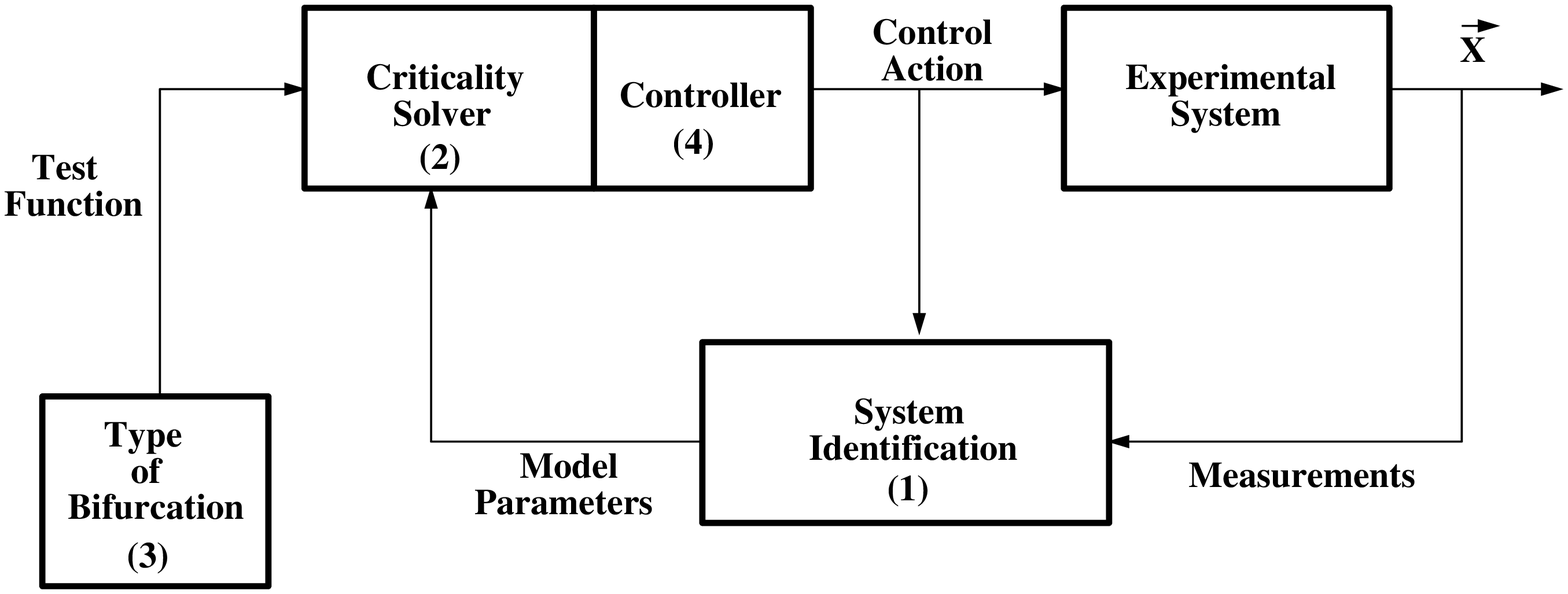,width=5.0in}}
\caption{Schematic diagram of the algorithm for the experimental
detection of bifurcations.}
\label{schemecontrol}
\end{figure}

The building blocks that constitute the bifurcation detection
protocol are schematically depicted in Figure \ref{schemecontrol}.
They involve four different modules: one that
identifies {\it on-line} an approximate model of the system, a second one
that uses the identified model to estimate the location of the
bifurcation point in phase$\times$parameter space; this second module
relies in the information provided by a test function (third
module) that contains the
conditions defining the type of bifurcation sought. 
The fourth and final module uses the estimated critical parameter value
(our estimate of the open-loop critical conditions) to devise a policy that
will bring and keep the system closer to open-loop criticality.
The repeated application of the algorithm, composed by these four
modules, drives the system to the true open-loop bifurcation point,
and holds it there.
In devising the algorithm, we assume
that at least one of the states of the experimental system can be
observed (measured), and that at least one operating parameter is available to
be manipulated (the bifurcation will be located with respect to this parameter).
Furthermore, we assume that the open-loop system is in the neighborhood of a
local bifurcation; thus its dynamics can be captured via a low-order
nonlinear model. 
Using such a low-order approximation to model the dynamics of the
system is motivated (and can be justified) by the existence 
of a normal form for the bifurcation.
The discrete-time model used in the system identification module of the
algorithm
((1) in Fig.~\ref{schemecontrol})
takes the form of a low order polynomial on the observed variables and the
parameter, motivated from Taylor series expansions in the neighborhood
of the bifurcation point, and the theory of Center Manifolds/Normal Forms.
The polynomial model is of the type:

\begin{eqnarray}
x_{i}(k+1) &\equiv& F_{i}[{{\bf x}}(k);p(k)] \nonumber \\
& = & a_{i} + \sum_{j=1}^n b_{i,j}x_j(k) +  c_{i}p(k) +
\sum_{j=1}^n \sum_{l=1}^n d_{i,j,l}x_j(k)x_l(k)\nonumber \\
&  & + \sum_{j=1}^n  e_{i,j}x_j(k)p(k)+
\sum_{j=1}^n \sum_{l=1}^n \sum_{m=1}^n f_{i,j,l,m}x_j(k)x_l(k)x_m(k)
\quad   (i = 1, \cdots,n)  \nonumber  
\end{eqnarray}

\noindent
where ${\bf x}$ represents the $n$-dimensional
vector of state variables, $p$ is the parameter, $a,b,c,d,e$ and $f$
are the polynomial coefficients, and $k$ is the iteration counter.
Note that the polynomial is linear in the parameter and contains cubic terms
on the state variables. The normal forms of
the most common bifurcations are linear
in the parameter, and the normal form of the
Hopf bifurcation has cubic terms in the state variables \cite{Wig}.
The polynomial can be simpler for certain specific types of bifurcation;
the form presented here is appropriate for the detection of
local codimension-1 bifurcations.
For the experimental system, we rely in Poincar\'e sections 
and/or reconstructions based on delayed measurements to form the vector 
of states.
The identification module uses a least-squares algorithm \cite{PTVF} to fit the
coefficients of the polynomial from experimental observations;
other algorithms, such as the projection method, can of course also be used. 
The model is updated at every step as new observations become available.
In order to prevent ill-conditioning of the least-squares algorithm during
the experimental run, a well-known problem in the
{\it on-line} identification of experimental systems,
the parameter signal must be persistently excited
(here, by addition of random noise) \cite{astrom,G&S}.

The second module of our algorithm ((2) in Fig.~\ref{schemecontrol})
uses the polynomial model identified
by the first module to estimate the location of the open-loop bifurcation.
This is a computational step that
involves the simultaneous solution of the steady state equations
along with a criticality condition appropriate for the bifurcation sought
((3) in Fig.~\ref{schemecontrol}).
This is accomplished using an algebraic equation solver (usually a contraction
mapping like a Newton-Raphson). 
For example, for a period-doubling bifurcation, 
which we detected in the R\"ossler circuit, 
the bifurcation is located at the point where one Floquet multiplier
(an eigenvalue of the linearization around the fixed point of the
open-loop Poincar\'e map model) is equal to -1. 
Thus, a suitable test function, for
a two-variable discrete-time system, will be $det[{\bf{I}}+{\bf{J}}]= 0$,
where $\bf{I}$ is the identity matrix and
${\bf{J}}=[\nabla_x{\bf F}[{\bf x};p]]$ is the Jacobian of the
identified model at the steady state $({\bf x},p)$.

In order to find the location of the critical 
state ($\{{\bf \hat{x}}_{cr},\hat{p}_{cr}\}$),
one must solve for $\{\bf{x},p\}$ such that (period doubling case):

\begin{eqnarray}
x_i-F_{i}[{\bf{x}};p]& = & 0 \quad   (i = 1, \cdots,n) \nonumber \\
det[\bf{I}+\bf{J}] & = & 0  \nonumber  
\end{eqnarray}

The final module ((4) in Fig.~\ref{schemecontrol})
uses the identified model and the estimated location of
the bifurcation ($\{{\bf \hat{x}}_{cr},\hat{p}_{cr}\}$) to find a parameter
variation protocol that will drive the system to the bifurcation point.
This critical step can be accomplished using several alternative ``feedback
laws". Previously \cite{ciss,prl},
we had illustrated the use of one of the simplest:
a policy that will bring the system to criticality in the
minimal possible number of steps (dead-beat policy).
In the runs presented below, however, we rely
on ``softer" policies in order
to overcome problems associated with the local character of the model
identified, the limited resolution of measurements and
the need for continuous excitation of the control signal associated with
the ill-conditioning of the least squares identification procedure.
In addition, the control policy should be possible to be
evaluated and implemented very fast (almost instantaneously) in real time.
This is because we have ignored in our experiment the delays
associated with I/O tasks as well as those associated with
the computation of the control action - we have effectively
assumed that computation and implementation of control action
is instantaneous. 
This avoids a number of complications and possible instabilities 
having to do with physical delays in the control loop.
These are
important issues, and will be taken up in future work; they are,
however, more ``implementation difficulties", and do not directly
pertain to the mathematical underpinnings of our procedure.

With these restrictions in mind, we explored a couple of 
plausible two parameter variation
protocols: the first policy attempts to minimize the distance from the
critical point in the next iteration, penalizing large
parameter variations; 
the second protocol is based on decreasing the distance to the critical
state (in phase space) in the next iteration by a fixed fraction.
In the notation used above, the first protocol results from:

\begin{equation}
\min_{p(k)} \sum_{i=1}^n (x_i(k+1)-\hat{x}_{i,cr})^2 + W (p(k)-\hat{p}_{cr})^2
\end{equation}

Note that, since the identified model (used to predict
${\bf x}(k+1)$) is linear in the parameter $p$, the solution of
this minimization problem renders a linear equation for $p$.  
The parameter $W$ is a penalty weight, which we usually set in the range [0.1,10].
Since the model form used is only valid locally,
close to the open-loop bifurcation, its predictions away from the
bifurcation are not accurate. 
In the presence
of modeling errors, a dead-beat policy may drive
the system farther away from the bifurcation instead of bringing it
closer. 
The introduction of the weight $W$ gives us a mechanism to 
moderate excessive control action, and thus, to possibly 
overcome the effect of 
such modeling errors away from the bifurcation point.
A more appropriate implementation of this policy
would seek only to penalize actions that bring the system
outside the range of validity of the model. 
That is, the penalty should only be applied if the prescribed control action
is larger that the estimated range of validity of the identified model.
We are currently exploring this issue.

The second policy is motivated by the form of transients approaching a
steady-state near a Hopf bifurcation. In such a scenario the dynamical
system will slowly converge to a marginally stable steady state by
(possibly fast, but slowly decaying) spiraling towards it. 
Thus, this second policy attempts to control the
speed of the approach by requiring that the distance from the critical
state location in the next iteration becomes a fraction of the {\it current}
distance from  criticality. In mathematical terms we solve for $p(k)$ such
that:

\begin{equation}
\sum_{i=1}^n (x_i(k+1)-\hat{x}_{i,cr})^2 = C^2 \sum_{i=1}^n
(x_i(k)-\hat{x}_{i,cr})^2
\end{equation}

\noindent
Here $C$ is the desired fractional reduction, in the next iteration,
of the distance from estimated open-loop criticality. 
We typically set $C$ in the range [0.9,0.95].
Since the model to predict $x_i(k+1)$ is linear in $p$, the solutions
of this equation are the roots of a quadratic equation that can be solved
explicitly. The root selected is the one closer to the current value of
the parameter.

Both control policies must be applied repeatedly in order to drive the
closed loop 
system to the point that, for the open-loop system,
was a marginally stable solution. Although the number
of iterations
to reach the bifurcation may be large (e.g. more than 100
iterations), we have found that, even with these
simple policies, the closed-loop
system performs well, finding the bifurcation for every experiment
independently of the initial condition used.

By using the identification and control methods in the manner described, we 
circumvent the long settling periods associated with open-loop
near-critical dynamics. As previously illustrated in \cite{prl}
through computational modeling, this
methodology is also capable of driving the system back to steady 
state after (mildly) overshooting the critical value of the
bifurcation parameter even for 
``hard" bifurcations such as turning points or subcritical Hopf bifurcations.
In the following section, we 
illustrate our approach by locating
period-doubling bifurcations of 
periodic orbits for the R\"ossler circuit device.

\begin{figure}[t]
\centerline{\psfig{file=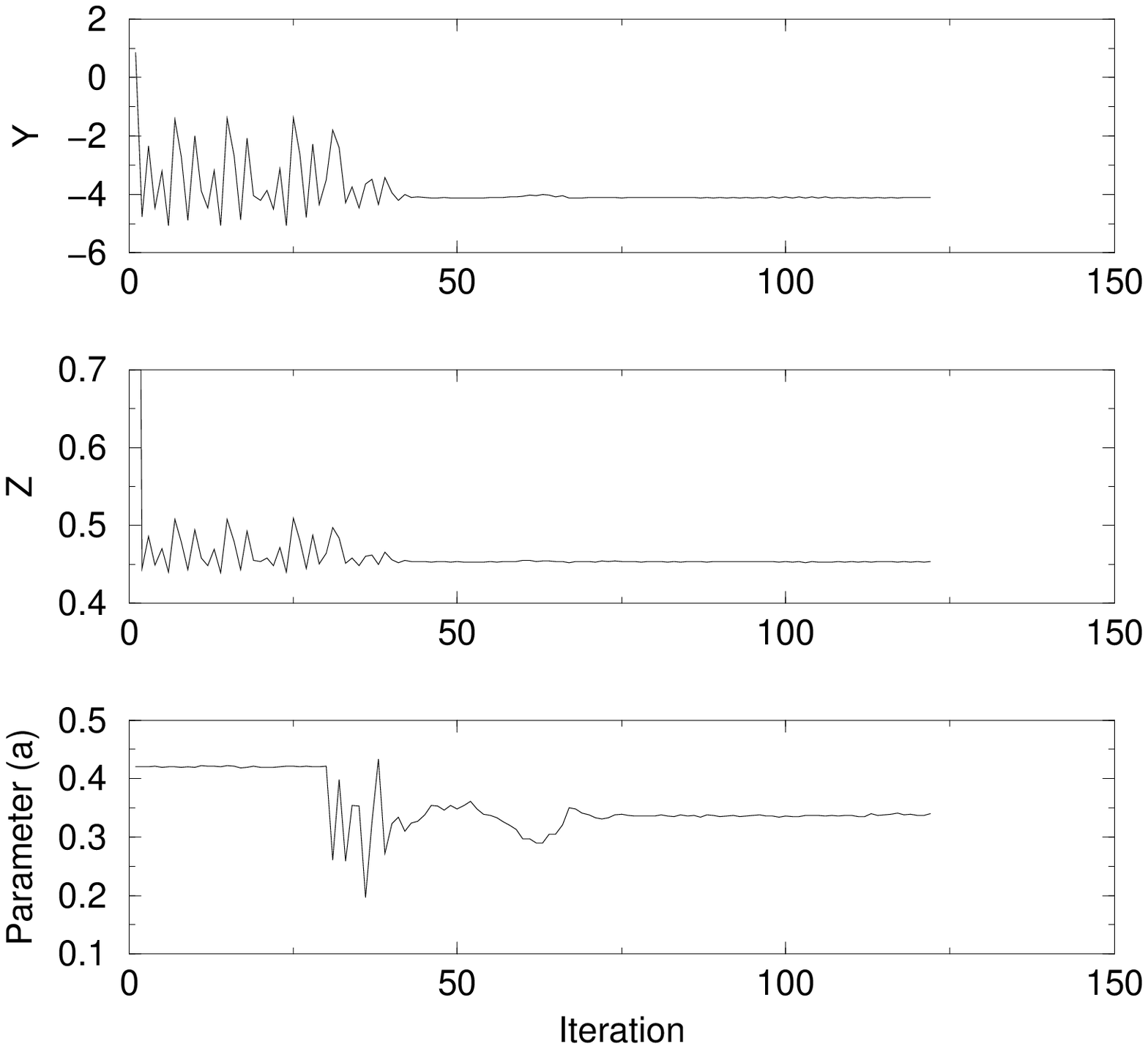,width=5in}}
\caption{Experimental results for the location of the period doubling
bifurcation of the period-1 limit cycle in the R\"ossler circuit. After
a short transient (50 steps) the initial model is properly identified and the
circuit is driven to the bifurcation point. The parameters $b$ and
$c$ were set at 2.0 and 4.0 respectively.}
\label{fig1}
\end{figure}

\section{{\bf Experimental Runs.}}

A period-doubling cascade of periodic orbits (or fixed points of
maps or Poincar\'e maps) is perhaps one of the
best understood routes to deterministic chaos. 
The R\"ossler system exhibits this
type of transitions at $b=2$ and $c=4$ for increasing values of the
remaining parameter $a$. The first of such period doubling bifurcations
is observed for $a\approx 0.33485$. It is precisely this bifurcation that
we set to locate using the protocol described above.

Poincar\'e sections provide a natural framework for observing/reporting
period doubling of periodic orbits.
The construction of our approximate models is performed on
such a (transversally intersecting the flow) Poincar\'e section.
Initially, we measure all three state variables from
the circuit $(x,y,z)$ and define the Poincar\'e section as
the plane $x=0.2$, crossed towards increasing values of $x$. 
For this Poincar\'e section
the critical point is located at $y=-4.1409$ and $z=0.44601$.
In the proposed protocol, the model to be identified involves two state
variables (the coordinates $(y,z)$ on the Poincar\'e section) and one parameter
($a$). Thus, our model will try to predict where the trajectory
intersects the plane at the next iteration ($y_{k+1},z_{k+1}$)
based on the current
location ($y_k,z_k$) and the value of the parameter ($a_k=p_k$). For
the experiment, the intersection point on the surface defining the
Poincar\'e surface
was obtained via fast (between 200 to 1000 Hz) sampling of all variables and
linear interpolation between the sampled points once the crossing with
the surface was detected.

\begin{figure}[t]
\centerline{\psfig{file=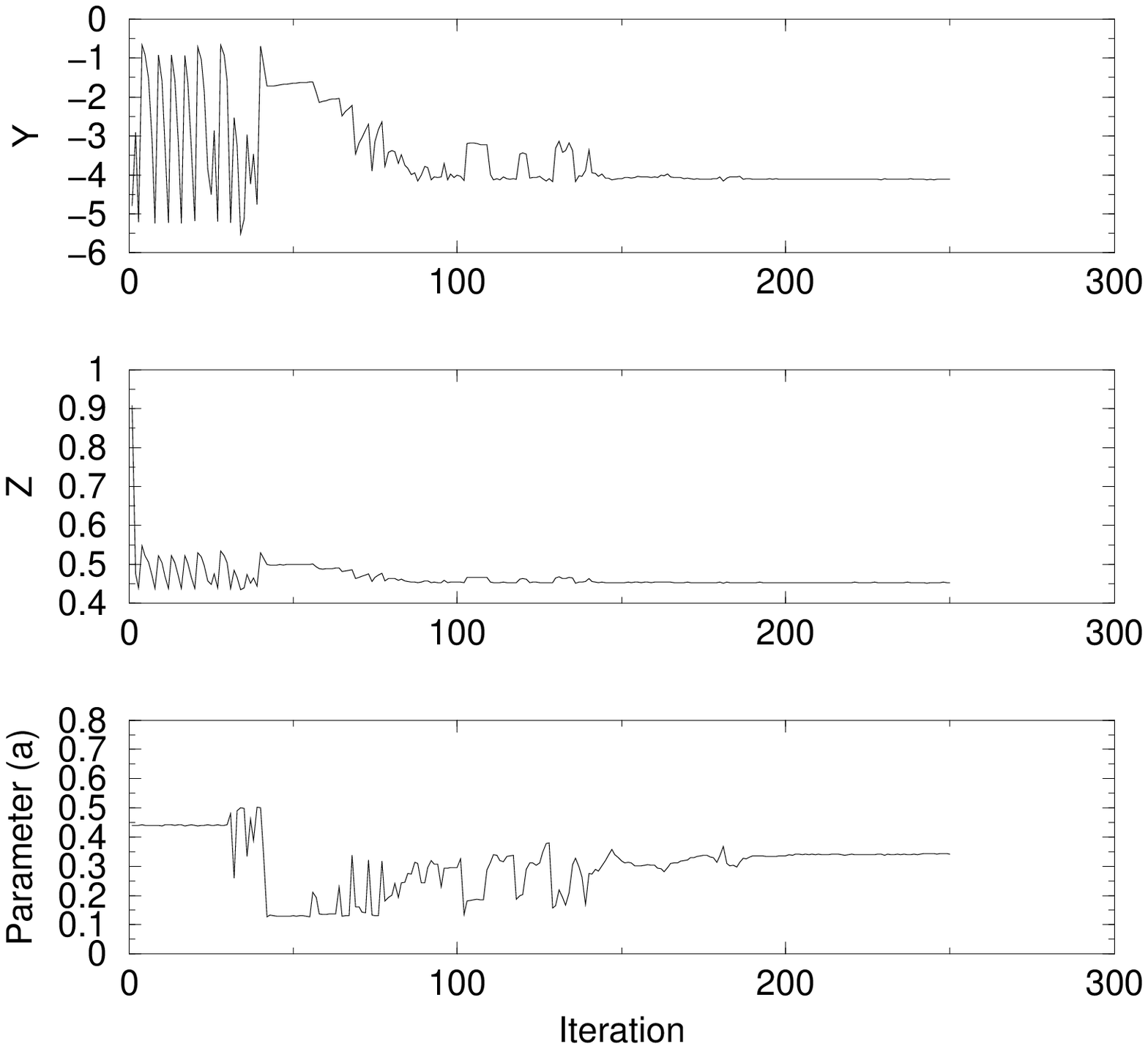,width=5in}}
\caption{Experimental run exhibiting the effect of the lower bound
for the control parameter. The parameter is initially ``trapped"
near the lower bound until a better model is identified thus restarting
the movement towards the bifurcation point. The
parameters $b$ and $c$ were set as indicated in Figure 4.}
\label{fig2}
\end{figure}

For the circuit, the period of the oscillations before the bifurcation
is approximately 0.62 s. In comparison with the R\"ossler ODE model, the
arbitrary units of time translate as seconds for the circuit device and the
arbitrary units of the state variables and
parameters as voltages measured in Volts.
Using a 16-bit data acquisition card, the measurement resolution for
the state variables is approximately 0.0006 volts and
0.00003 volts for the parameter $a$
which has a smaller range of variation (voltage). 
As we mentioned earlier,
in order to successfully identify a varying {\it on-line} model, one should
apply persistent excitation to the actuator (the parameter signal).
This is done in the
form of random noise with maximum amplitude of 0.003 volts. The excitation
signal prevents the system from reaching stationary states for which the
measurement will be effectively constant.
Such measurements will render the matrix used in the identification 
computations severely ill-conditioned, and the approximate model
will deteriorate. 
With persistent excitation signals the system
will be always ``on the move" allowing the collection of information-rich data
resulting in more accurate models. 
Since, however the final goal is to
drive the system to the open-loop bifurcation, very large excitation signals are 
undesirable because they work against the control actions implemented. 
The excitation level selected
represents a compromise between these competing effects. Depending on the
system under consideration, the magnitude of the excitation signal
should be adjusted.
In our experience with several different simulated systems, we have found that
the amplitude of the excitation signal must be at least one order of magnitude
larger than the resolution available in the measurement.

The use of this relatively large excitation signal combined with the soft
control policies described previously constrains the selection 
of the appropriate convergence criterion. 
The approach to the
critical state may be slow, possibly also ``erratic" due to the
effect of the excitation; on the other hand, the system may become trapped
and the limited
resolution may translate in biased models and incorrect criticality
predictions. 

\begin{figure}[t]
\centerline{\psfig{file=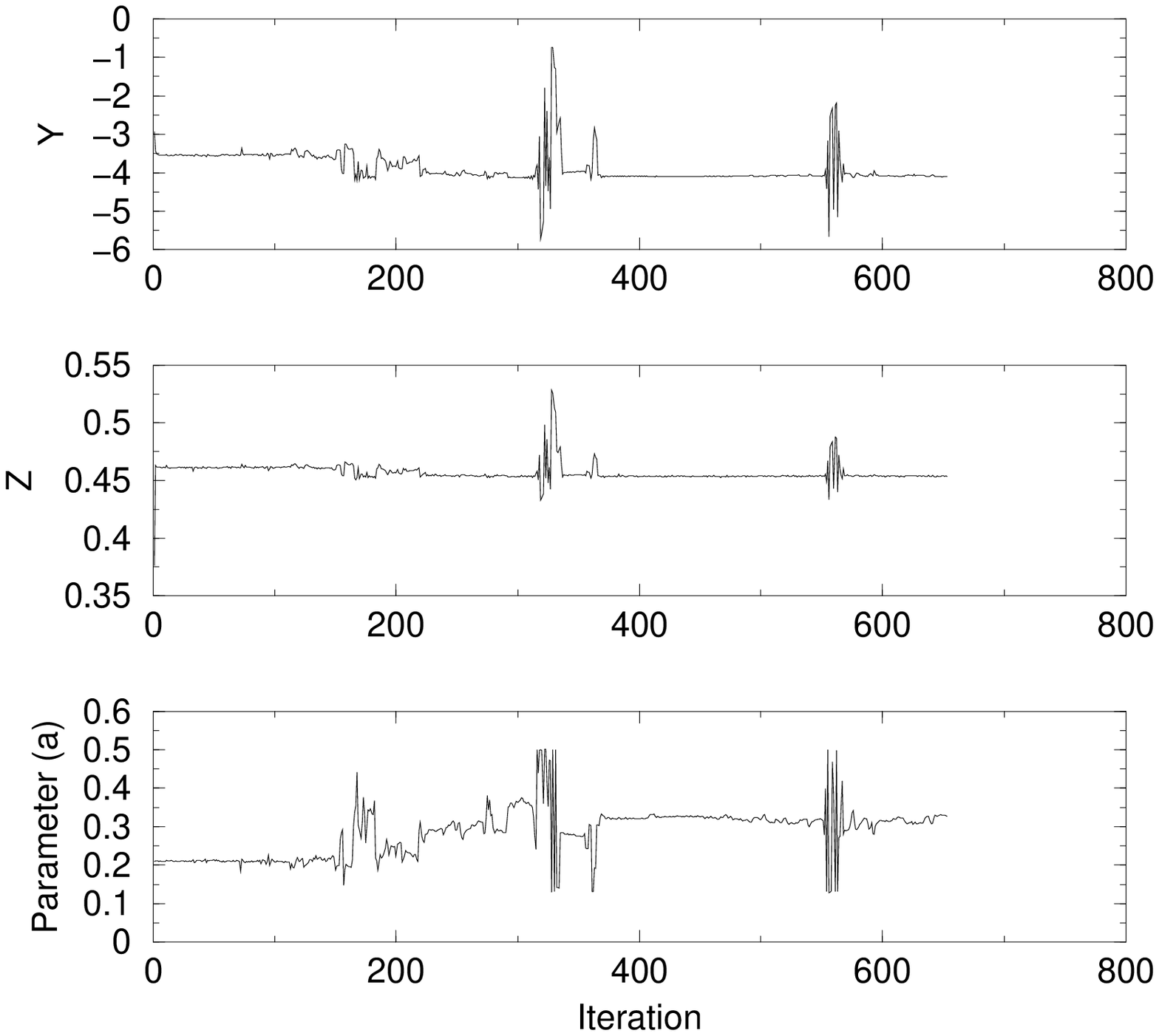,width=5in}}
\caption{Experimental run exhibiting bursting.
The adaptive bifurcation driver settles for a while in a point near the
critical coordinates, without achieving convergence,
until near-singularity of the matrix used
for the identification produces a sudden burst that drives the
circuit far away. The
parameters $b$ and $c$ were set as indicated in Figure 4.}
\label{fig3}
\end{figure}

In order to prevent the latter occurrence, we opted for a
two-level convergence criterion: (1) we require that the norm of
the deviation from criticality in the extended
phase$\times$parameter space is sufficiently small (i.e. less than
0.008 volts) for a relatively small
number of consecutive iterations (typically 5); and (2) after complying
with the first criterion, we restart the system identification
procedure by gathering enough data
to construct a ``fresh" model, using the excitation signal
without control, and then restart the feedback control until the first criterion
is again satisfied. 
Overall convergence is declared only
if two consecutive restarts reach the same point within the tolerance
indicated above. The first criterion does not excessively burden the
controller performance, while the second seeks to prevent declaring
convergence at a possibly unconverged,
incorrect prediction of the location of the bifurcation.
The tolerance to declare convergence (0.008 volts in our case)
should clearly be larger than the set
maximum amplitude of the random noise (excitation
signal, 0.003 volts); a reasonable rule-of-thumb is to
set this tolerance half-an-order of magnitude above the maximum
amplitude of the random noise.
We note, however, that the achievable accuracy in locating the bifurcation
is limited, not because of the convergence criterion, but most
importantly because of the excitation signal, model sensitivity and,
ultimately, limited resolution in the measurements.

With the feedback protocols implemented, we observed, during our runs,
that large parameter variations (``control actions") 
arose frequently.  Since such
variations may drive the system outside the range physically accessible to
the circuit device (i.e. negative parameter values), 
or outside the range for
which sustained oscillations are observed (thus, preventing the
trajectories from intersecting the plane used in the definition of
the Poincar\'e section) we opted for establishing hard bounds
(``saturation") for the values that the parameter can take. 
If the control
action prescribed goes beyond these bounds, the parameter is
set to the value of the bound that has been surpassed (the actuator saturates).
For the run at $b=2$ and $c=4$, the minimum value allowed for $a$ was 0.12
and the maximum $0.5$. Below
$a\approx 0.125$ the only attractor of the system is a steady state,
while above $0.5$ the chaotic attractor observed in the range [0.396,0.5]
disappears via global bifurcations and the state variables of the
circuit tend to saturate. 

With the hardware available to us, the evaluation of the protocol,
including identification of an updated model, location of criticality
and evaluation of the control policy, involved between 30 and 60 milliseconds,
depending on the number of evaluations required for some of the subroutines
such as the algebraic solver to find criticality. The effect of such ``large"
delay (about one tenth of the period of oscillation in the worst case)
was often detrimental to the performance of the protocol, preventing the accurate
location of the bifurcation. In order to overcome this situation, we 
modified the evaluation order of the subroutines. That is, we first
calculate and implement a control policy using one-iteration-old models,
then we use the collected data to update the model and prediction of
criticality for the next iteration.
This strategy was successful, since the evaluation of the
control policies proposed require only the solution of linear or quadratic
equations and thus take little time (less than 3 ms), and the slow approach
to the critical state makes consecutive models very similar most of the
time (depending on the conditioning of the matrix involved in the
identification of the model).

Figure ~\ref{fig1} illustrates the performance of the protocol.
The parameter was initially set at $a=0.42$. At this parameter
setting, the open-loop circuit exhibits a chaotic attractor. 
The experiment is left to evolve in the vicinity
of this parameter value for about 30 iterations, while data are gathered
to construct an initial model. 
After approximately 70 iterations, the system
has been driven to the vicinity of the bifurcation. The last 50 iterations,
for which the response of the circuit is practically constant, represents
the period needed to declare convergence with one restart as described above.
For this run the critical state was predicted at $y=-4.1127$, $z=0.4537$
and $a=0.3389$.
This compares favorably with the location of the bifurcation
for the R\"ossler ODE system. 
As we indicated before, we anticipated that the
estimation of the bifurcation location will not be very accurate
because of the persistent excitation signal and soft control policies. 
A more representative result should be reported
after gathering statistics over a number of runs. On average, over 15 runs,
each run with different initial condition and initial parameter
value, the
bifurcation is predicted to be located at $a=0.3309$ with a standard deviation
of $0.0082$ volts, with the critical state located at $y=-4.0964$
and $z=0.4552$ and standard deviations of $0.0134$ and $0.0014$ volts
respectively. 
The deviation of this average is well within the range
allowed for declaring convergence in our runs and the standard deviation
commensurate with the excitation signal and convergence criteria.
We note that all our runs did converge to the critical state
albeit occasionally at a larger number of iterations than for Figure~\ref{fig1}.

\begin{figure}[t]
\centerline{\psfig{file=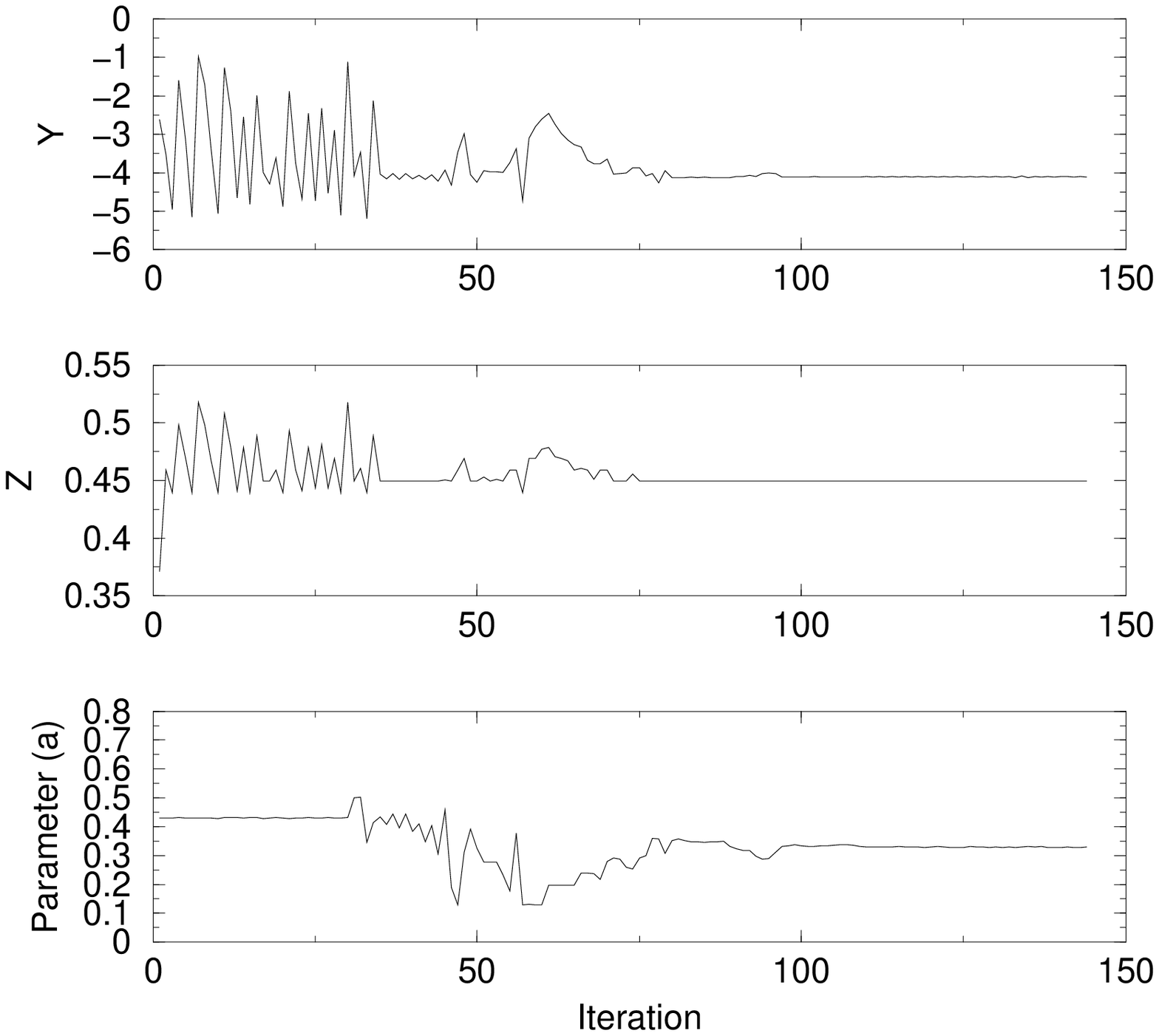,width=5in}}
\caption{Experimental run with lower resolution. For a resolution based
on an 11 bit A/D converter, the bifurcation is also properly
located. The parameters $b$ and $c$ were set as indicated in Figure 4.}
\label{fig4}
\end{figure}

Figure~\ref{fig2} presents another run that illustrates the role of
hard, saturation bounds limiting the values that the parameter can take
during the search for criticality. 
After approximately 40 steps, an incorrect criticality prediction translates 
into a drastic control action.
Since the value prescribed for the parameter would be below the preset lower
bound, the parameter is set to the bound. The performance of the controller
deteriorates and the system is trapped near the lower bound for about 40
iterations before restarting its movement towards the critical
state.

Figure~\ref{fig3} shows an example of the phenomenon
commonly referred to as ``bursting".
The system starts its gradual approach toward criticality, but since
the coordinates of the points in the trajectory do not change much, the
matrix used for the identification becomes progressively more ill-conditioned,
eventually rendering a ``bad" model that, in turn, dictates a large parameter
excursion (``burst"). The trajectory to criticality exhibits, in this
example, a couple of instances of such bursting on its route to the
eventual correct detection of the critical fixed point.

For those runs initialized at parameter values lower than the bifurcation value,
as the one illustrated in
Figure~\ref{fig3}, the convergence to the critical point was
consistently declared for slightly lower values of the
parameter than the true critical parameter value. 
In contrast, when the initialization took place at conditions 
for which the open loop system exhibited apparently chaotic oscillations (i.e.
Figure~\ref{fig1}), the collection of information-rich data allowed
the accurate estimation of the critical fixed point location early on, 
resulting in faster convergence towards the bifurcation.

One potential limitation of the application of this strategy may be the
available measurement resolution. The examples presented above rely on a 
16 bit A/D card. However, as illustrated in Figure~\ref{fig4}, the
bifurcation was successfully located also for lower resolution, equivalent
to that provided by a standard 11 bit A/D converter (i.e. one order
of magnitude less in resolution for the parameter and state variables).

\begin{figure}[t]
\centerline{\psfig{file=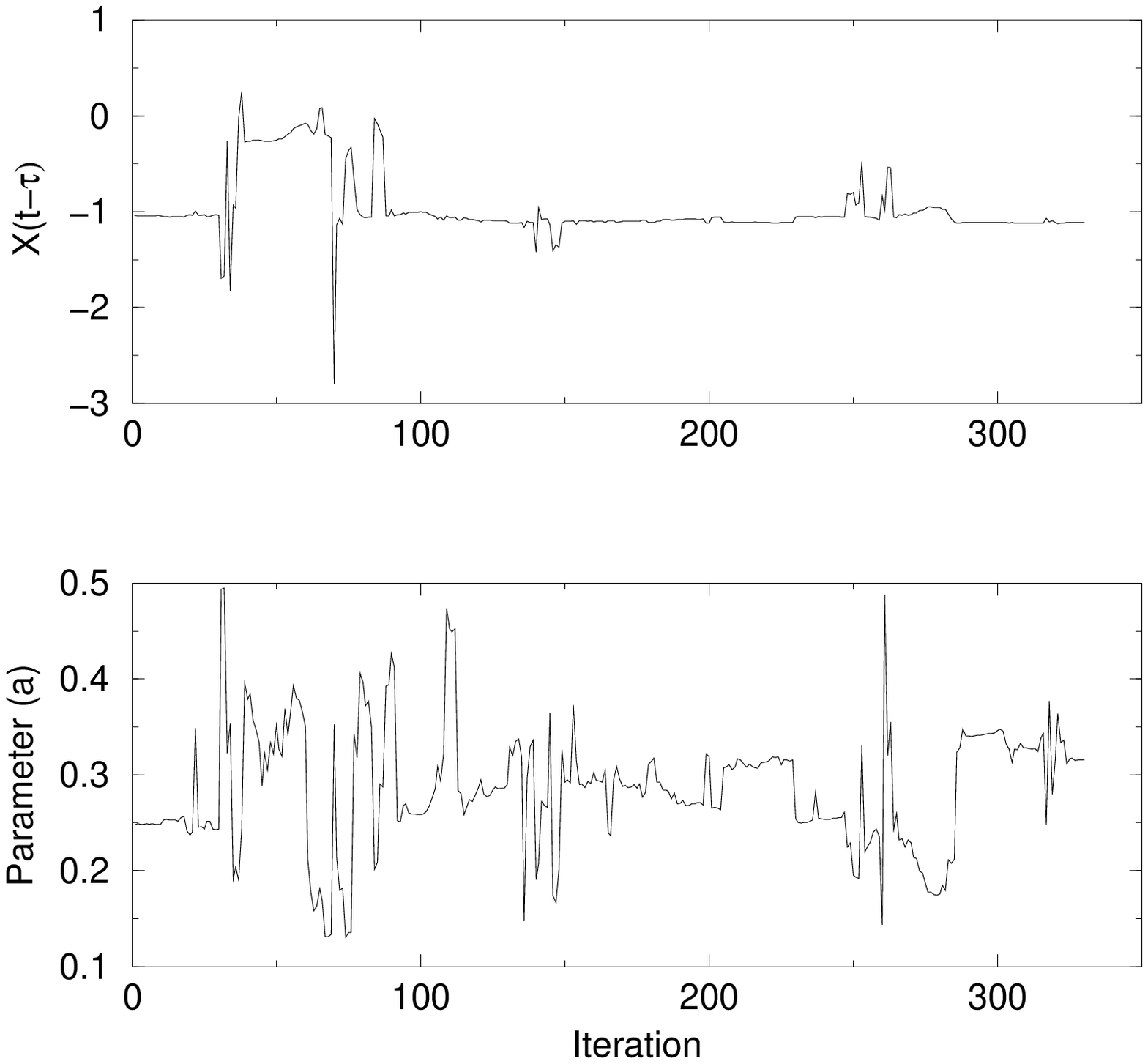,width=5in}}
\caption{Experimental run using  measurements of only one
variable and a delay-based reconstruction of
the phase-space as the basis to take the Poincar\'e section.
The delay was set to 40 ms, two delays were used and
the Poincar\'e section was set for the current value of the
measured variable (x) equal to 0.2.
The parameters $b$ and $c$ were set as indicated in Figure 4.}
\label{fig5}
\end{figure}

Although the examples discussed until this
point involve measurements of all the states of
the system, it is also possible to use only a single measurement time series.
Figure~\ref{fig5} illustrates the performance of the protocol for
the location of the same bifurcation using only measurements of the
variable $x$ and relying on delayed measurements to reconstruct the phase space.
The delay ($\tau$) for the reconstruction was set to 40 ms. In practice one
could seek to find an optimal value of the delay using, for example,
mutual information measures; this was not done here.
The Poincar\'e section was set as before using the plane $x(t)=0.2$.
The critical state is located for the R\"ossler system at
$x(t-\tau)=-1.1701$ and $x(t-2\tau)=-2.2020$. On average over 5 runs,
the bifurcation is predicted at $a=0.3235$ for
$x(t-\tau)=-1.1143$ and $x(t-2\tau)=-2.1320$, with standard deviations
of $0.0080$, $0.0042$ and $0.0010$ volts respectively. The trajectory to reach
the critical state, however, exhibits a larger number of
instances of bursting and saturation of the parameter and thus requires
a larger number of iterations as illustrated
in Figure~\ref{fig5}. This deterioration in performance of the protocol
can be attributed to the increased sensitivity of the identification
resulting from the use of a sub-optimal delay and the smaller effective
resolution of the sampled variable.

\begin{figure}[t]
\centerline{\psfig{file=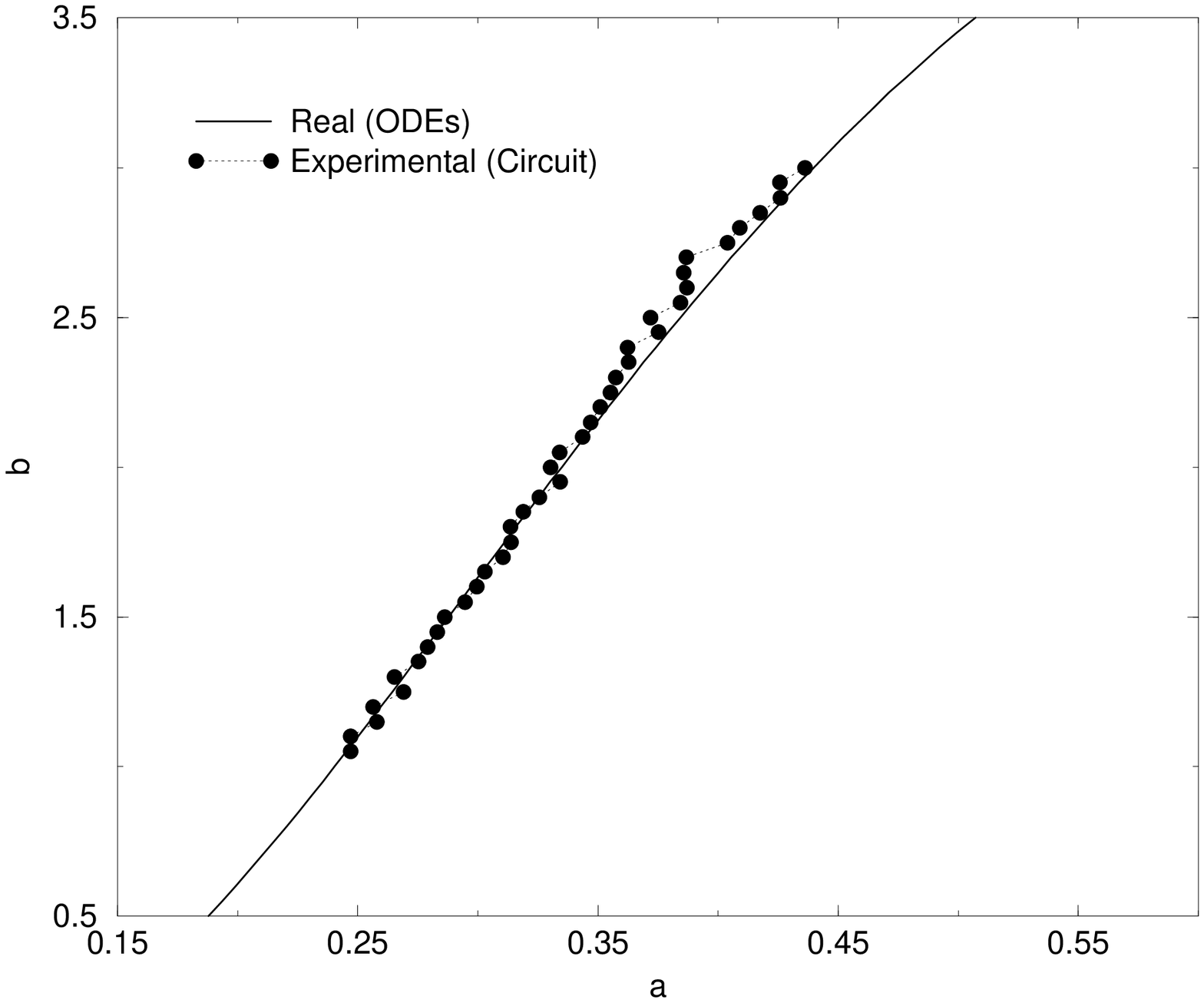,width=5in,angle=270}}
\caption{Comparison of the experimentally
obtained two parameter boundary and the numerically calculated
from the R\"ossler System. The parameter $b$ was changed gradually
and the bifurcation point obtained from the previous value used as
starting point to locate the new bifurcation. The experimental
results are averaged over 5 runs, three runs using a variation
for $b$ decreasing and two for $b$ increasing.
The parameter $c$ was set to 4.0.}
\label{fig6}
\end{figure}

\begin{figure}[t]
\centerline{\psfig{file=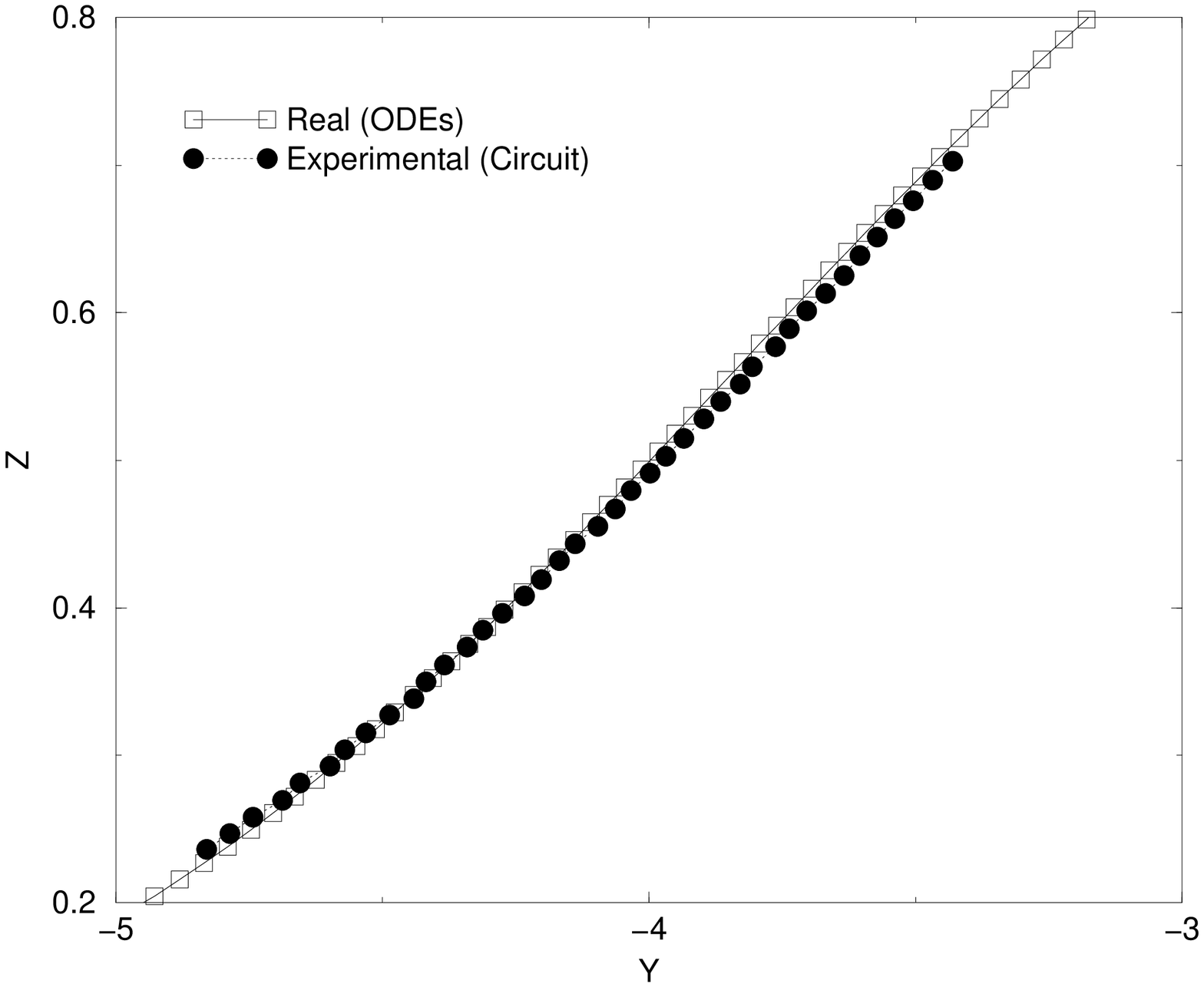,width=5in,angle=270}}
\caption{Comparison of the experimentally
converged location of the marginally stable fixed points
and the numerically calculated ones 
from the R\"ossler system. The parameter $b$ was changed as indicated
in the previous figure and $c$ was set to 4.0.}
\label{fig7}
\end{figure}

A final test of the capabilities of the proposed protocol for the
detection of bifurcations comes by attempting to draw, experimentally,
a two-parameter bifurcation diagram. 
That is, we attempt to trace automatically
a segment of our codimension-1 bifurcation boundary in two parameters. 
We vary the (second) parameter $b$ in the range [1,3]. 
Starting at $b=3.0$, we locate the value of $a$ for which the 
bifurcation is observed and then
decrease $b$ by a fixed amount (0.05 volts). 
Figure~\ref{fig6}
presents the comparison of these runs for our circuit device,
and the computed boundary for the R\"ossler ODE system. 
The results presented are the
average over five sweeps, three
calculated decreasing $b$ and two increasing it. The previously found
bifurcation point was used as the starting point for the new search
with a different value of $b$.
Figure~\ref{fig7} presents the comparison of the location
of the (open loop critical) fixed points for the Poincar\'e section.
The boundaries for the R\"ossler ODE system were
calculated using AUTO \cite{Doe}.
As can be seen in Figures~\ref{fig6} and ~\ref{fig7}
the segment is accurately located, as are
the critical states. For both plots, standard deviations have been
omitted; they are comparable to the previously indicated
values for the single parameter runs.

\section{{\bf Conclusions}}

We have presented a successful implementation of
an algorithm for efficiently driving an experimental system to 
(what at open loop would have been) criticality.
The control protocol was applied to an experiment in
the form of an electronic circuit device that
implemented the dynamics of the R\"ossler system.
The proposed algorithm is based on adaptive system identification, 
separation of time scales close to criticality, and on the
exploitation of the identified local model to search for the
criticality conditions; the estimation
of open-loop criticality is used to devise a control policy that drives the
closed-loop system towards the bifurcation.
The protocol seeks, in a sense, to mimic the procedure used in the numerical
construction of bifurcations from mathematical models:
the critical state of the open-loop system becomes a {\it stable} steady state of 
the {\it augmented} system in (phase) $\times$ (operating parameter) space.
We illustrated the capabilities of the algorithm in the location of
period doubling bifurcations of periodic orbits.
The bifurcation was successfully located even in the presence of limited
resolution.
We also illustrated the applicability of this approach when only
one state variable can be measured by reconstructing the phase space
using delayed measurements.
Our final illustration involved the tracing of a segment of the
two parameter bifurcation boundary.

Our selection of an electronic circuit device with R\"ossler dynamics,
to demonstrate the experimental feasibility of the protocol for
the automatic detection of bifurcations, was intended to provide
a testing ground for assessing the performance of the protocol.
The circuit device provides us with a highly reproducible and easily 
manipulated experiment with a well charted dynamical behavior.
We chose the period doubling bifurcations of the oscillations
of the circuit as our representative codimension-1 bifurcation.
The low-dimensionality of the model used crucially depends on the 
time-scale gap between the long time scale associated with
the critical eigenvalue(s) and the time scale of the ``next least
stable" eigenvalue. 
It is important that all operations performed in our system 
(I/O as well as software computations) are fast compared to 
this ``next least slow" time scale.
Since the return time for our Poincar\'e map was roughly 600 ms,
our relatively short (approximately 40 ms) 
action delay could be considered ``approximately instantaneous".
An alternative that would speed up the protocol evaluation is the use of
recursive least squares techniques \cite{astrom,G&S}
that may reduce significantly
the time involved in the identification of a local model; however, this
alternative resulted in more sensitive and prone-to-burst trajectories,
even though we attempted several different strategies in the assignment of
forgetting factors for the identification.
Other alternatives, such as the use of projection algorithms \cite{G&S},
are also feasible, but were not investigated.

Perhaps the most limiting aspect for the widespread application of the
proposed protocol involves the high sensitivity of the strategy
as a result of the combination of the control policies, excitation
signal and on-line identification of a model with only local validity.
Such sensitivity translates in erratic and slow motion towards the
bifurcation, frequent bursts, trapping of trajectories near hard bounds
and ill-conditioning of the identification
process.

We are currently working on posing the entire problem (system, identification
and control) as an augmented dynamical system, for which the existence 
of an overall converged state can be demonstrated, and its stability quantified.
Such an approach could also allow us to explore issues related to the
performance in the presence of multiple nearby bifurcations of the same type.

Extensions of the algorithm are straightforward: higher codimension
bifurcations can be detected by solving the corresponding (more than one)
criticality conditions for the (open-loop) marginally stable steady state and the
(more than one) critical parameters.
In principle one might even be able to accommodate unstable, nonbifurcating
parts of the spectrum on the right-hand plane.
Several important elements can be improved upon:
(a) Formulation of the criticality conditions in a less
naive and computationally more efficient manner, along lines of modern
numerical bifurcation theory (e.g. \cite{Doe}
and references therein); (b) Usage of more sophisticated
(e.g. receding horizon) feedback control laws \cite{rawlings};
and (c) Improvement of the type of local model used.
While in this problem we used polynomial models based on Taylor-series
expansions and normal forms, it is obvious that other
types of nonlinear dynamic models (e.g. models based on artificial neural
networks) can also be used in this context; we, among many others, have 
demonstrated that such models are capable of  accurately predicting
the bifurcation structure of a nonlinear system when trained with time series
data \cite{greece,bulsari}.

While we are continuing theoretical and computational work along the directions 
described above, we are also implementing this approach in the
study and characterization of the bifurcations of electrochemical
systems \cite{Wol}.

It may be interesting to note that this approach can also be coupled with
large scale scientific
simulators, for which the relevant long term dynamics are low-dimensional,
in order to chart their bifurcation scenario \cite{copeland} (as
opposed to writing a bifurcation algorithm for them from scratch).
The strategy of seeking and stabilizing a fixed point with one additional
condition (here, open-loop criticality), falls in the same category as
extremum seeking controllers \cite{krstic}; controller developments along
these lines would find direct application in our protocols, and vice versa.
We believe that an approach like the one described here can eventually become a
standard real-time tool which will assist experimentalists in
efficiently detecting bifurcations and
tracing operating diagrams; such a tool could drastically
enhance the experimental characterization of nonlinear dynamics
in multiparameter space.

\section{{\bf Acknowledgements}}

We are grateful to Professor Ertl for his interest in 
and generous support of this work.
We thank  Heinz Junkes and Georg
Heyne from the FHI technical staff for technical support.
We gratefully acknowledge the support of CONACYT and COSNET (RRM),
NATO (IGK, KK), and AFOSR (IGK) for partial funding of this research. 
IGK and RRM gratefully acknowledge
the support from the Alexander von Humboldt Foundation.

\end{document}